\documentclass[sigconf, nonacm, screen]{acmart}
\usepackage{graphicx}
\usepackage{wrapfig}
\usepackage{ifthen}
\newboolean{techreport}
\usepackage{caption}
\usepackage{multirow}
\usepackage{subcaption}
\PassOptionsToPackage{hyphens}{url}
\usepackage{hyperref}
\usepackage{amsmath}
\usepackage{tikz}
\usepackage[]{footmisc}
\usepackage{graphicx}
\usepackage{pifont}
\usepackage{amsmath,amstext}
\let\oldnl\nl
\newcommand{\nonl}{\renewcommand{\nl}{\let\nl\oldnl}}
\usepackage{tikz}
\usepackage{algorithm}
\usepackage{algpseudocode}
\usepackage{amsmath}
\usepackage{xspace}

\usepackage{arydshln}
\usepackage{threeparttable}

\definecolor{codegreen}{rgb}{0,0.6,0}
\definecolor{codegray}{rgb}{0.5,0.5,0.5}
\definecolor{codepurple}{HTML}{C42043}
\definecolor{backcolour}{rgb}{1,1,1}

\newcommand\numberstyle[1]{%
    \footnotesize
    \color{codegray}%
    \ttfamily
    \ifnum#1<10 0\fi#1 |%
}

\usepackage{enumitem}
\renewcommand\footnotetextcopyrightpermission[1]{} 

\begin{document}

\title{DiStash:  A Disaggregated Multi-Stash Transactional Key-Value Store}

\author{Yiming Gao}
\affiliation{%
  \institution{University of Southern California}
  \city{Los Angeles, CA}
  \country{USA}
}
\email{gaoyiming@usc.edu}

\author{Hieu Nguyen}
\affiliation{%
  \institution{eBay Inc.}
  \streetaddress{2025 Hamilton Avenue}
  \city{San Jose}
  \state{CA}
  \country{USA}
  \postcode{95125}
}
\email{hieunguyen@ebay.com}

\author{Jun Li}
\affiliation{%
  \institution{eBay Inc.}
  \streetaddress{2025 Hamilton Avenue}
  \city{San Jose}
  \state{CA}
  \country{USA}
  \postcode{95125}
}
\email{junli5@ebay.com}

\author{Shahram Ghandeharizadeh}
\affiliation{%
  \institution{University of Southern California}
  \city{Los Angeles, CA}
  \country{USA}
}
\email{shahram@usc.edu}

\begin{abstract}
A stash is a storage medium such as Dynamic Random Access Memory (DRAM), Solid State Disk (SSD), Hard Disk Drive (HDD), or Non-Volatile Memory (NVM).  This paper presents a disaggregated transactional key-value (KV) store, DiStash, that governs KVs across pools of stash types.  
It enables an application to use a single transaction to read and write different copies of one or more key-value pairs across the different pools of stashes.  It simplifies the application logic by (a) preventing undesirable race conditions that may cause copies of data across different stash pools to become inconsistent and/or (b) failures that may result in loss of key-value pairs. A configuration of DiStash may use a pool of stashes as either ephemeral or durable storage.  
The application dictates whether the content of its participating stashes are inclusive (replicated) or exclusive (tiered).  We implement a DiStash by extending FoundationDB.
We quantify the tradeoffs with its design decisions using microbenchmarks and eBay’s production workload.  We open source our implementation at \url{https://github.com/ebay-USC/DiStash}.
\end{abstract}


\maketitle

\section{Introduction}
Advances in computing, networking, and mass storage devices have introduced disaggregation of storage.
Dynamic Random Access Memory (DRAM)~\cite{lenfant1977fast,cenker1979fault}, Solid State Disk (SSD)~\cite{dirik2009performance,chen2016internal}, Hard Disk Drive (HDD)~\cite{gray1987,multizone1996,patterson1988case}, and alternative forms of Non-Volatile Memory (NVM)~\cite{patil2025nvm,wang2025boosting} including NVDIMM~\cite{lee2025boosting,clouzet2024h2m,lee2020nvdimm} are a type of stash that offer different speed, storage capacity, failure characteristics, and price points~\cite{config2018,config2018arXiv}.
To provide online access to a large data set, an application may construct pools of stash types.
It may either duplicate or migrate data across the different pools with the objective to maximize the number of references to the pool with the fastest stash.
This principle is the cornerstone of cache-augmented database management systems~\cite{writeback2019,mercury2012,ebay2023,scalingmemcache2013,onehoparXiv}, load balancing techniques using a front-end cache~\cite{smallcache,netcache2017}, and hierarchical storage structures~\cite{flashstore2010,hierarchical1995,nhc2021}.
A challenge of such systems is how to maintain copies of data consistent across different stash pools.
The current practice of using different transactional storage managers with different pools may result in undesirable race conditions that cause copies of the data to become inconsistent.
This means the value of a data item may reflect different values produced by the execution of different transactions~\cite{iq2014,consistency2010}.

\begin{figure}[!ht]
    \centering
   \includegraphics[width=1.0\linewidth]{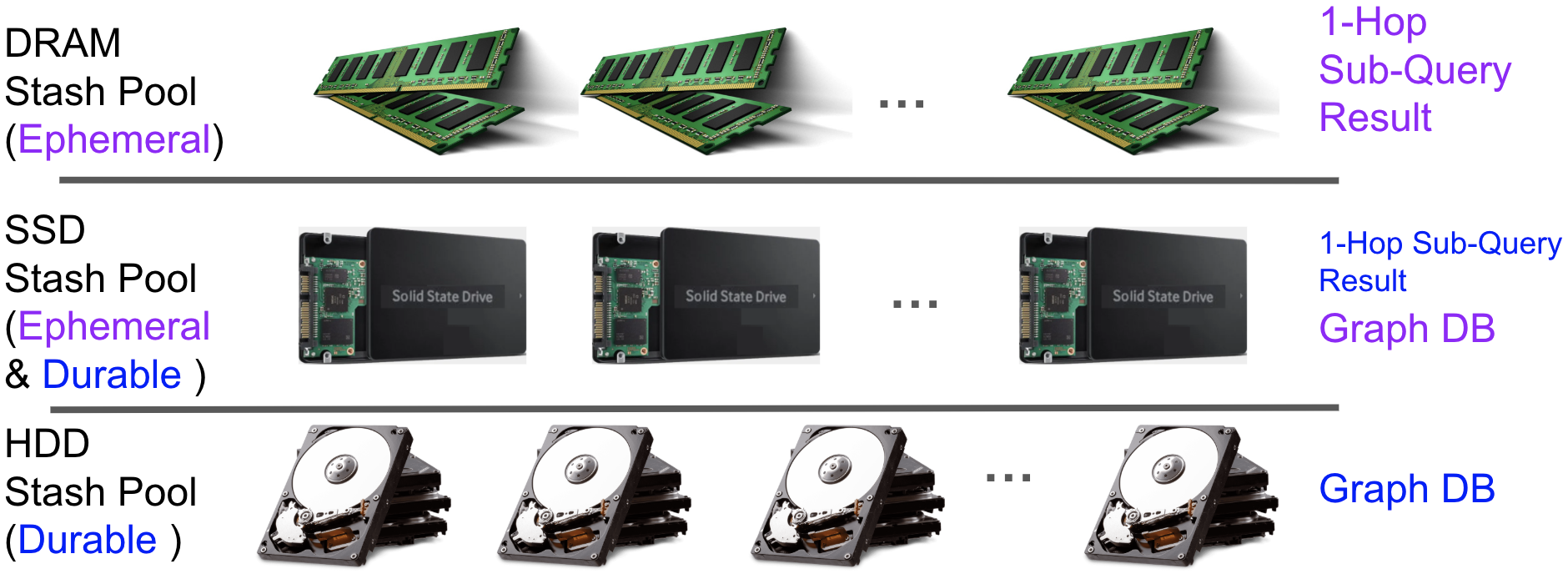}
    \caption{DiStash with three pools of stashes:  DRAM, SSD, and HDD.
    The pool of DRAM stashes is for ephemeral storage of 1-hop sub-query results~\cite{onehoparXiv}.
    The pool of HDD stashes is for durable storage of graph data.
    The pool of SSD stashes is ephemeral storage for graph data and durable storage for 1-hop sub-query results.
    A graph query processor may lookup sub-query results in either DRAM or SSD, and process queries using either SSD or HDD.
    }
    \label{fig:stashpools}
\end{figure}

DiStash\footnote{DiStash is a play on the word “destash” which means to declutter or switch locations.} is a disaggregated software architecture that builds upon the FoundationDB~\cite{zhou2021foundationdb,zhou2023foundationdb}.
It manages data across different stash pools using one transaction.
The stash pools may be scattered across different data centers.
A configuration of DiStash may specify a different degree of replication for data in a stash pool.
This is to tolerate failures of stashes in that pool.
An application may maintain a copy of the data on different stash types as illustrated in Figure~\ref{fig:stashpools}.

\begin{figure}[!ht]
    \centering
   \includegraphics[width=0.5\linewidth]{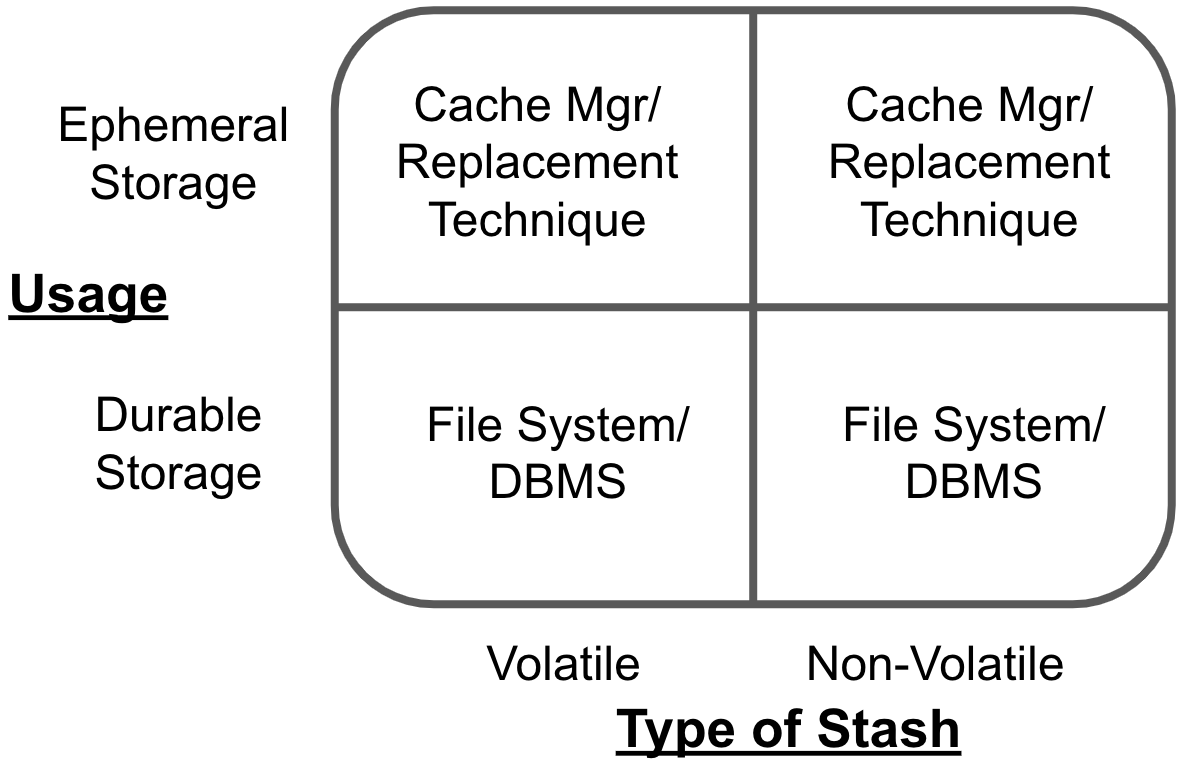}
    \caption{A taxonomy of stash types and their usage.
    }
    \label{fig:usage}
\end{figure}

A stash type may be volatile or non-volatile, see the x-axis of Figure~\ref{fig:usage}.
While a volatile stash such as DRAM loses its data after a power recycle, a non-volatile stash preserves its data. 
A DiStash deployment supports the designation of a stash type's operational mode as ephemeral or durable, see the y-axis of Figure~\ref{fig:usage}.
Ephemeral may provide the illusion of an infinite amount of storage space by maintaining a subset of data and evicting others. 
Examples include a cache manager such as memcached~\cite{officialmemcached}, Kosar~\cite{kosar2014}, or Redis~\cite{carlson2013redis}.
Durable maintains data permanently and has a fixed storage capacity.
Once it acknowledges storage of a key-value pair, the key-value pair persists.
Examples include POSIX-compliant file systems such as EXT2 and database management systems, DBMSs, such as SQLite.

Ephemeral and durable behave differently once their storage capacity is exhausted. 
Should the application attempt to insert additional data, ephemeral may delete perviously inserted key-value pairs to free space for the new key-value pair.
Durable returns an error, e.g., ENOSPC (Error No SPaCe) with 
EXT2 and SQLITE\_FULL with 
SQLite.





{\bf Contributions} of this paper include:
\begin{enumerate}
    \item DiStash, a transactional storage manager that supports pools of stashes each with a different number of instances and degree of replication.  The stashes may be spread across multiple data centers.  Section~\ref{sec:distash}.
    \item A DiStash transaction may read and write data from different pools of stashes.  Section~\ref{sec:why1}.
    \item An implementation of DiStash using FoundationDB~\cite{zhou2021foundationdb,zhou2023foundationdb}.  It may be deployed across multiple data centers.  Section~\ref{sec:impl}.
    \item An evaluation of two DiStash deployments across 3 data centers using eBay production workload.  Section~\ref{sec:eval}.
    \item We open source our software at \url{https://www.github.com/ebay-USC/DiStash}.  
\end{enumerate}
The rest of this paper is organized as follows. 
Section~\ref{sec:motivate} motivates DiStash by presenting several use cases of it.
Subsequently, we describe each of the above contributions in turn. 
Section~\ref{sec:related} presents related work.
Brief conclusions and future research directions are presented in Section~\ref{sec:conc}.

\section{Motivation:  Applications of DiStash}\label{sec:motivate}
DiStash has wide applicability described by the following use cases.

\subsection{Cache-Augmented Data Stores}
DiStash may be used to implement cache augmented data stores~\cite{onehoparXiv,scalingmemcache2013} in support of read dominated workloads such as social networking~\cite{scalingmemcache2013} or fraud detection~\cite{ebay2023}.
It may consist of 2 stash pools.
SSD for durable storage of data used for query processing.
DRAM for ephemeral storage of query results for their fast lookup with repeat queries.
Conceptually, the data on the different stashes is inclusive (replicated).  However, the physical representation of data is exclusive (tiered) because the result of queries is different than the data used to process those queries.
With eBay's production workload, a cache-augmented graph database lowers 95$^{th}$ and 99$^{th}$ percentile of query response times by at least 2x and 1.63x, respectively~\cite{onehoparXiv}.

Section~\ref{sec:cadbms} presents cache-augmented data stores in greater detail.  Briefly, an update may be required to modify data on SSD and the impacted query results in DRAM.
DiStash enables the application to perform both in one transaction.
Section~\ref{sec:why1} highlights the significance of using one transaction.
It shows with multiple transactions to manage data across different pools of stashes, undesirable race conditions may cause copies of data across these stash pools to reflect different values produced by different transactions.
DiStash frees the developer to focus on the application requirement instead of identifying the different race conditions and writing software to prevent them.
Writing software is expensive because it must be designed, developed, tested, and maintained.
DiStash eliminates this expense.




\subsection{Load Balancing}
A challenge of using a pool of $n$ SSD stashes is uneven distribution of load across individual stashes~\cite{smallcache,netcache2017}.
While data placement schemes help balance an aggregate load, they do not balance the dynamic load.
Hotspots may occur with a transient skewed access pattern. 
One way to balance load is to use a small front-end stash (a cache) for frequently accessed data items~\cite{smallcache}.
An implementation of this solution has been shown to improve throughput by 3-10x~\cite{netcache2017}.
The front-end stash must be fast enough to keep the $n$ slow SSD stashes busy.  
While a skewed access pattern negatively impacts the load balancing of the ($n$) slow stashes, it simultaneously enhances the effectiveness of the fast frontend stash in two key ways. Firstly, it achieves a higher hit rate. Secondly, it results in a more balanced distribution of uncached data across the ($n$) slow stashes.
It has been shown that the cache is required to only store O($n~log~n$) data items where $n$ is the total number of slow stashes, i.e., SSDs~\cite{smallcache}.

The resulting two-tier deployment must cope with unpredictable shifts in the query workload, e.g., flash crowds~\cite{smallcache}.
A miss for a data item in the front-end stash may populate the cache with a fixed probability.
A concurrent write for the same data item may implement one of the write policies such as write-around (Figure~\ref{fig:write-around}), write-through (Figure~\ref{fig:write-through}), and write-back~\cite{mercury2012,writeback2019}.
This write may race with the population of cache by the miss, causing the miss to populate the front-end stash with stale data.
See discussion of Figure~\ref{fig:race} in Section~\ref{sec:why1}.

DiStash prevents the undesirable race condition by enabling the load balancing solution to use one transaction to update copies of the data in the fast front-end stash and one or more of the slow stashes.

\subsection{Hierarchical Data Storage}
A storage hierarchy~\cite{hierarchical1995,nhc2021} may consist of $h$ pools of stashes, $S_0$ to $S_{h-1}$.
These stash pools may be ordered with $S_0$ as the fastest and $S_{h-1}$ as the slowest\footnote{It is possible for two or more adjacent stash pools to provide approximately the same performance~\cite{nhc2021}.}.
Typically $S_0$ is the most expensive and provides the least amount of storage compared with the other layers of the hierarchy.
The KV pairs migrate across the stashes based on their popularity and the working set of an application.
Ideally, the application working set should reside in the fastest stash, $S_0$~\cite{flashstore2010}.

Given a read request for a key, if it is not found in Stash $S_0$ then the system must look for it in the next stash $S_1$.
If it is not found in $S_1$ then it must search for it in $S_2$.
This process repeats until either the referenced key is found or all the $h$ stashes have been searched.
Bloom filters may be used to expedite identifying a missing key in a stash~\cite{rocksdb2017,novalsm2021}.
A key found in stash $S_i$,$~i\geq1$, might be moved to Stash $S_0$ based on the anticipation that it will be referenced again.
The application may delete its copy from Stash $S_i$ to ensure data across different stashes is exclusive. 

A write to a key-value pair may be written to $S_0$.
As $S_0$ becomes full, the application may migrate the infrequently referenced data to $S_1$.
This process may repeat for the other intermediate stashes with $S_{h-1}$ providing the highest density\footnote{Typically $S_{h-1}$ provides the lowest cost per byte and slowest service time.}~\cite{flashstore2010}.

Flashstore~\cite{flashstore2010} is an example key-value store, consisting of three stash pools: DRAM, SSD, and HDD.
It has been shown to outperform BerkeleyDB with two stash pools by providing 85x cost efficiency
(ops/sec/dollar)~\cite{flashstore2010}. 
Using our terminology, its durable storage is the HDD where the key-values reside.
It has two ephemeral storage. One for SSD to materialize the working set of an application.
A second for the DRAM to maintain the least recently referenced key-value pairs, a write buffer, and a hash index that points to the physical location of the keys in the SSD.
The potential race conditions that may result in data inconsistency are well documented, see Section 4.6 of~\cite{flashstore2010}.
In addition, this study identifies consistency with multiple stash as non-trivial,
differing it to future work, see Section 4.7 of~\cite{flashstore2010}.

DiStash may be used to implement a system similar to Flashstore~\cite{flashstore2010}.  This implementation preserves consistency of data while allowing for multiple instances of a stash type.
A stash pool in the hierarchy may scale independently and have a different degree of replication.
See Section~\ref{sec:distash}.

\subsection{Hybrids}
It is possible to have a hybrid of the above possibilities. 
This is shown in Figure~\ref{fig:stashpools}.
The DRAM, SSD, and HDD is a hierarchy similar to Flashstore~\cite{flashstore2010}.
The DRAM and SSD is a cache augmented solution similar to~\cite{onehoparXiv}.  
This figure shows the SSD pool as both ephemeral and durable storage.
It is ephemeral for graph data and durable for 1-hop sub-query results.
The configuration of DiStash that implements the hybrid maintains 4 stash pools:
one DRAM, one HDD, and two SSD.
The two SSD stash pools are independent:
One for durable storage of 1-hop sub-query results and a second for ephemeral storage of graph data.

A graph query consisting of one or more 1-hop sub-queries may be processed as follows.
The query processor looks up the result of each 1-hop sub-query in DRAM.  If it is not found there, the query processor looks it up in the SSD durable storage.  If it is not found there then the query processor executes the query using the instance of the graph database on the SSD ephemeral storage.  If the referenced vertices and edges by the query are missing from SSD then they are staged from HDD to SSD.
The query processor may compute the result of the missing 1-hop sub-query results to populate the DRAM stash.
As the DRAM stash becomes full, it writes its entries to SSD.


\section{DiStash}\label{sec:distash}
A DiStash is a disaggregated storage manager for migrating and replicating copies of data across pools of stashes in a transactional manner.
An application may use it to implement a cache augmented solution.
Below, we describe how the concept of a transaction prevents undesirable race conditions that result in loss of data or inconsistencies across different replicas of a data item.

Without loss of generality, for the rest of this paper, the term stash and stash pool is used synonymously. 

\subsection{Cache Augmented Data Store}\label{sec:cadbms}
An application may configure DiStash with SSD as durable storage for use by a query processing engine and DRAM as ephemeral storage for the result of queries.
When a query is issued, the application looks up the result of the query in the cache.
If found then the application has observed a cache hit, returning the cache entry as the result of the query.
Otherwise, a cache miss, the application executes the query, computes its results, generates a cache entry, and inserts it in the cache for future lookup~\cite{scalingmemcache2013}.

\begin{figure}[!ht]
    \centering
   \includegraphics[width=0.5\linewidth]{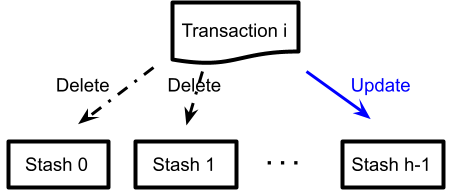}
    \caption{Write-Around.
    }
    \label{fig:write-around}
\end{figure}

In the presence of writes to the data in the durable storage, the application must maintain the cache entries in the ephemeral storage consistent.
It may use different write policies~\cite{jouppi1993cache,writeback2019}.
Figure~\ref{fig:write-around} shows the {\em write-around} policy with a hierarchy of $h$ stash pools.
It requires the application to issue a transaction that deletes all copies of a cache entry and updates the original data.
Copies of the cache entries typically reside on the fastest stashes with the original data residing on the slowest stash. 
With two stash pools, $h$=2, the copy of query results in DRAM, i.e., the fast stash, is deleted while the original data on the SSD, i.e., the slowest stash, is updated.

\begin{figure}[!ht]
    \centering
   \includegraphics[width=\linewidth]{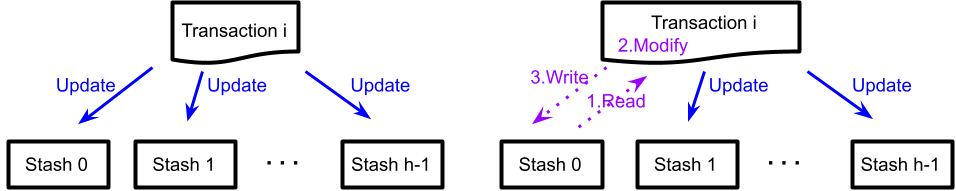}
    \caption{Write-Through.
    }
    \label{fig:write-through}
\end{figure}

Figure~\ref{fig:write-through} shows the {\em write-through} policy.  It requires the application to issue a transaction to update all copies of a cache entry and the original data across all $h$ stash pools.
The update to a stash pool might be in the form of an SQL update command, a memcached append command, or Read-Modify-Write (RMW).
With the latter, the transaction may read the impacted data from a stash, modify its value, and write it back.
Figure~\ref{fig:write-through} shows the RMW against one stash pool.
However, it may be performed across 2 or more and possibly all stash pools.
DiStash enables an application to use the write-around and write-through policies in different transactions concurrently. 

\begin{figure*}[ht]
    \centering
    \begin{subfigure}[b]{0.45\textwidth}
        \centering
        \includegraphics[width=\textwidth]{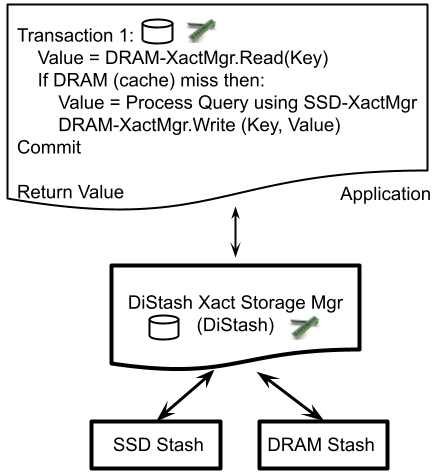}
        \caption{Read Transaction.}
        \label{fig:read}
    \end{subfigure}
    \hfill
    \begin{subfigure}[b]{0.35\textwidth}
        \centering
        \includegraphics[width=\textwidth]{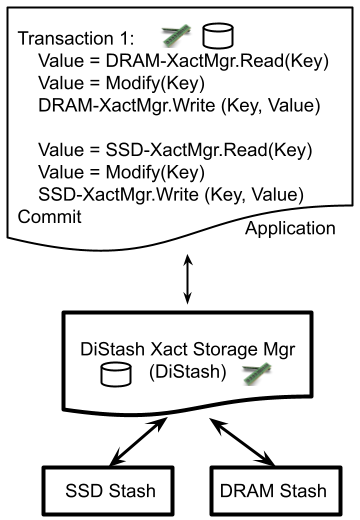}
        \caption{Write Transaction.}
        \label{fig:write}
    \end{subfigure}

    \caption{Read and write transactions with DiStash.}
    \label{fig:distashreadwrite}
\end{figure*}
Figure~\ref{fig:distashreadwrite} shows an implementation of a read and a write with DiStash for two stashes:  DRAM and SSD.
The transaction that reads data from DRAM becomes a read-write when it does not find its data in DRAM, see Figure~\ref{fig:read}.
The write implements the write-through policy by updating both the SSD and DRAM stashes in one transaction.
Both assume each stash represents its data as key-value pairs.
A key concept is one transaction that either reads or writes data from both stashes.

\subsection{Why One Transaction?}\label{sec:why1}
Without the concept of one transaction that implements ACID semantics across the same copy of data (either logical or physical) across all stashes, an implementation may suffer from race conditions that result in the loss of latest write.
We illustrate this with a deployment that consists of two different transactional storage managers:  
one manages the DRAM stash while the other manages the SSD stash.
This causes the application to implement the alternative policies by issuing different transactions to each stash.

\begin{figure*}[ht]
    \centering
    \begin{subfigure}[b]{0.425\textwidth}
        \centering
        \includegraphics[width=\textwidth]{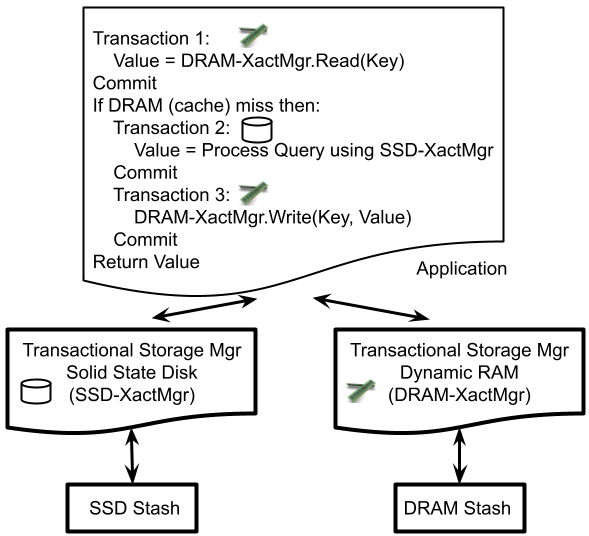}
        \caption{Read Transactions.}
        \label{fig:2smread}
    \end{subfigure}
    \hfill
    \begin{subfigure}[b]{0.45\textwidth}
        \centering
        \includegraphics[width=\textwidth]{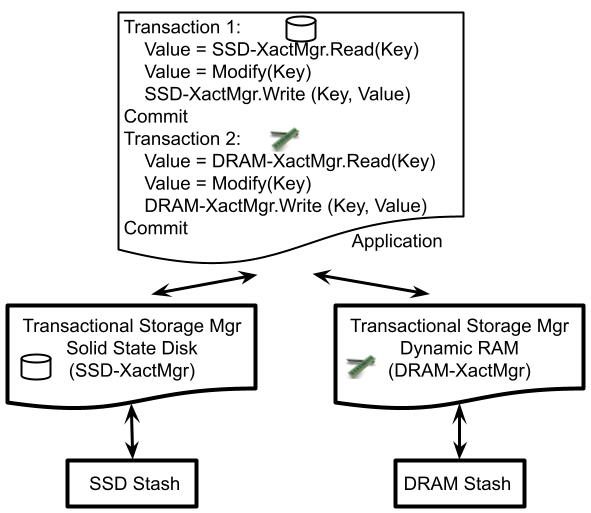}
        \caption{Write Transactions.}
        \label{fig:2smwrite}
    \end{subfigure}

    \caption{Read and write transactions with different stashes being managed by different storage managers.}
    \label{fig:2sm}
\end{figure*}

Figure~\ref{fig:2smread} shows the potential transactions issued by a read to the DRAM stash.
When the referenced key is found, it consists of one transaction.
However, with a miss, it consists of three transactions.  
The second transaction references the SSD storage manager to compute the result of the query.
The third transaction populates the DRAM with the result of the query.

Figure~\ref{fig:2smwrite} shows the write with two storage managers.  It consists of two independent transactions.  One per storage manager (that manages a different stash) to implement the write-through policy.

\begin{figure*}[!ht]
    \centering
   \includegraphics[width=\linewidth]{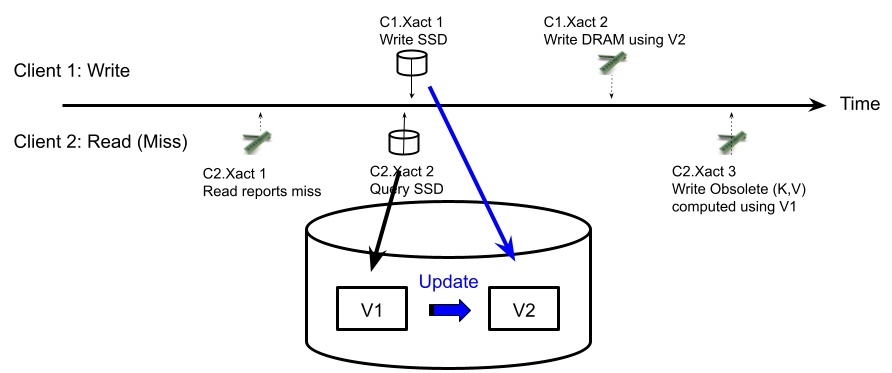}
    \caption{An undesirable race condition results in inconsistent data across DRAM and SSD.
    }
    \label{fig:race}
\end{figure*}

The two storage manager solution suffers from undesirable race conditions that may cause the different stashes to reflect inconsistent values. 
One such race condition is shown in Figure~\ref{fig:race}.
Client 1 issues a write that updates the SSD and DRAM stashes in turn per Figure~\ref{fig:2sm}.b.
Client 2 issues a read that observes a cache miss.
Hence, it issues a second transaction that reads data from the SSD stash.
This transaction overlaps with the SSD write of Client 1.
The underlying storage manager may use multi-version concurrency control (MVCC), enabling Client 2 to read an obsolete value of data.
If Transaction 3 of Client 2 executes after Transaction 2 of Client 1, 
the value written to DRAM does not reflect the value produced by the latest transaction that wrote SSD\footnote{Unless V1=V2.}.
A subsequent read using DRAM does not observe the value produced by the latest transaction.
This violates the isolation and durability properties of transactions.

Figure~\ref{fig:race} highlights the fact that while the transactions issued to each storage manager are serializable, transactions across the two storage managers are not serializable.
This causes copies of the same data item on two or more stashes to reflect inconsistent values.
DiStash addresses this by using one storage manager and one transaction to read and write from different stashes, see Figure~\ref{fig:distashreadwrite}.
This forces the transactions by Client 1 and 2 to be serialized:
Client 1's transaction followed by 2's transaction, or
Client 2's transaction followed by 1's transaction.
Should the operations overlap as shown in Figure~\ref{fig:race}, DiStash will abort one of the two transactions, forcing its client to retry its transaction.  
This ensures copies of a data item across different stashes to have consistent values.
This preserves the isolation and durability properties of transactions.

\section{An Implementation}\label{sec:impl}
\begin{figure*}[!ht]
   \includegraphics[width=0.9\linewidth]{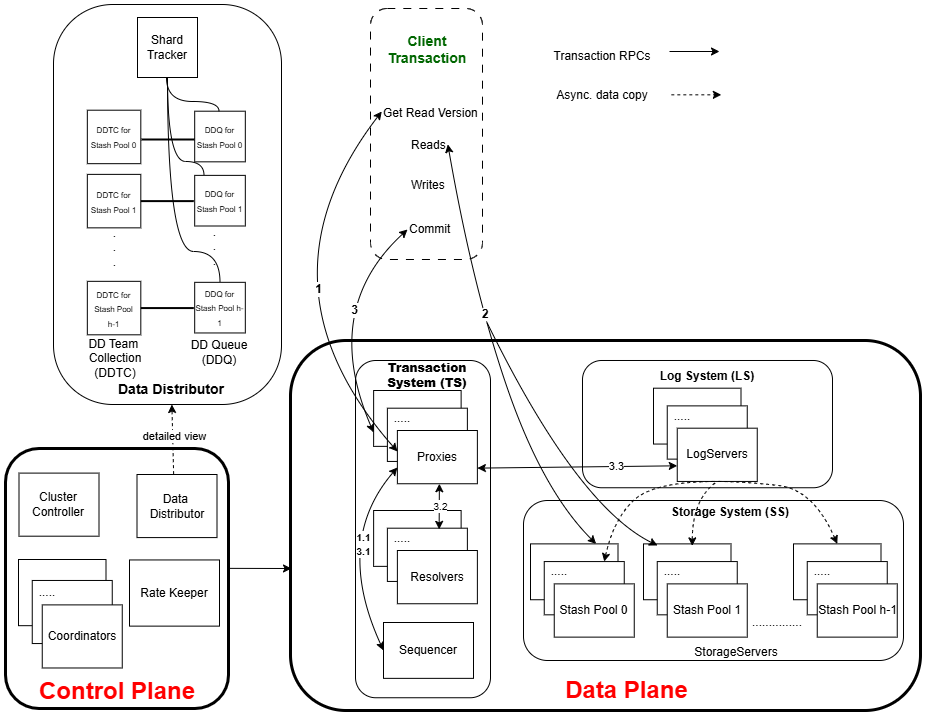}
    \caption{An implementation of DiStash. See~\cite{zhou2023foundationdb} for definition of arrows.}\label{fig:arch}
\end{figure*}



We extended FoundationDB to implement DiStash, see Figure~\ref{fig:arch}:
\begin{itemize}
    \item Support $h$ pools of stashes.
    \item Each stash pool may have a different degree of replication.
    \item One log server for all $h$ stash pools.
    \item A transaction may reference key-value pairs that reside across all $h$ stash pools.
    \item Balancing of load across the individual stashes that constitute a pool.
    \item Support for both ephemeral and durable storage using volatile and non-volatile stash pools.  See Figure~\ref{fig:usage}.
    \item Recovery of a non-volatile stash is different than a volatile stash.  With the latter, the recovery process uses a replica to maintain hit rate.
    \item One may extend a volatile stash with non-volatile storage for local logging, enabling the volatile stash to recover its data from short lived failures faster than 60 seconds.  The threshold 60 seconds is adjustable.
\end{itemize}

Key changes to FoundationDB include:
\begin{itemize}
    \item A Data Distributor (DD) team collection and queue for each stash pool.  The queue facilitates re-location of shards for load balancing across the stashes in a pool.
    \item A new Storage Server (fdbserver/KeyValueStoreCache.actor.cpp) that extends FDB's existing in-memory implementation with the LRU replacement technique in support of DRAM ephemeral storage.
    \item Extensions of the configuration file to specify the $h$ stashe pools in the system and whether each is ephemeral or durable.
    \item Initialization of the FDB to read from the configuration files.  This serves two purposes.  First, initialize DD for each stash pool.  Second, initialize the storage servers for each stash pool with the pre-specified degree of replication.  
    \item A different prefix is used for each stash pool.  This prefix is specified in the configuration files.
    The DD is aware of this prefix and this information is fetched by the client.
    \item We changed the Data Distributor Team Collection (DDTC) to maintain a different degree of replication for each stash pool.
    We also changed the initialization of the DD to verify the specified degree of replication for a stash satisfies the current deployment.
    If this is not satisfied then DD constructs replicas to meet the requirements of the deployments.
\end{itemize}
One may use an existing implementation of SQLite or RocksDB with a durable stash.
A future research activity is to implement an SSD optimized ephemeral storage similar to Flashstore.

DiStash partitions data across the stashes by assigning a unique prefix to a stash pool.
These prefixes are specified in a configuration file.
The Data Distributor (DD) directs keys with a prefix to its unique stash pool.
It consists of $2h+1$ processes:  one shard tracker process (DDST), $h$ Team Collection (DDTC) processes, and $h$ Queue (DDQ) processes\footnote{The DD of the original FDB consisted of 3 (2+1) processes, one of each of shard tracker, DDTC, and DDQ.  In our implementation, there is one extra process that is not shown.  It is called the Shard forwarder.  It is introduced to simplify the implementation.}.
The DD Shard Tracker ensures the data is partitioned across the stashes consistent with the prefix boundaries defined for each stash pool.   
A DDTC is responsible for grouping the machines and servers within its corresponding stash pool to construct replica groups.
A DDQ is responsible for migrating logical partitions, i.e., ranges, of data across the nodes of each stash pool by migrating ranges across its stashes.
The DDST may construct new ranges by splitting existing ranges and merging existing ranges into larger ones.



\section{Performance Evaluation}\label{sec:eval}
We evaluate DiStash using both microbenchmarks and eBay's production knowledge graph.  This evaluation considers normal and failure modes of DiStah.
The failure mode consists of epochs.  Each epoch starts with the failure of $x$ stashes and ends with the insertion of $x$ new stashes\footnote{$x$=1 with the microbenchmark and 2 with the eBay knowledge graph.}.

The microbenchmark quantifies the performance of DiStash with different stash types.
It focuses on data ingest by writing a fixed amount of data to either a DRAM stash or a SSD stash.

It shows how DiStash's use of MVCC~\cite{zhou2021foundationdb} results in a lower rate of data ingest with the SSD stashes.
With failures, the microbenchmark evaluates a deployment consisting of 
10 DRAM stashes 
across two data centers, 5 per data center.
The third data center maintains the log records using SSD.  
We use $x$=1 stash with each epoch in failure mode.

eBay's knowledge graph facilitates search using a graph representation of data.  It is a read dominated workload with the database refreshed periodically, approximately every 2 weeks.  It generates a persistent database consisting of final query result for a subset of query templates that require a low latency.

We use a trace driven evaluation of the production workload to compare two system configurations, 1Stash and 2Stash.
1Stash consists of a durable store using 40 SSDs for both graph data and query result in each of two data centers.
2Stash separates durable store for graph data using 40 SSDs from ephemeral store for query results using 5 DRAMs in each of two data centers.  

The size of the production graph database is approximately 2.1 Terabytes.
It is replicated 3 times across the 40 SSDs.
The size of the query result is approximately 12.6 Gigabytes.
It is replicated 3 times across 40 SSDs with 1Stash and 5 DRAMs with 2Stash.
A third data center maintains the log records using SSD.

The eBay workload consists of 5 query templates.
Three of these query templates are cached and constitute approximately 45\% of the workload. 
We considered two workloads:  Light and Moderate. 
Light uses 100 pods each with 10 threads to generate the workload.
Moderate uses 200 pods each with 10 threads to generate the workload.
In failure mode, $x$=2 stash with each epoch.
With 2Stash, the failure epochs apply to the DRAM stash pool only.

\subsection{Normal Mode of Operation}
This section presents results from DiStash normal mode of operation using our microbenchmark and eBay production workload in turn.  We start with a summary of the results.

\subsubsection{Summary of results:}
Our microbenchmark shows writing to DRAM is faster and more predictable when compared with either SSD or HDD.
DiStash's use of MVCC~\cite{zhou2021foundationdb} causes the rate of data ingest to fluctuate with SSD and HDD.
With the eBay workload, we consider the latency of the overall workload and the fraction that benefits from the query results, i.e., cache hits.
2Stash improves the 95$^{th}$ and 99$^{th}$ overall latencies by 10\% when compared with 1Stash.
With a low system load, the latency of the cache hits with 1Stash is superior to 2Stash.  This observation disappears with a moderate system load.

\subsubsection{Microbenchmark}
\begin{figure}[!ht]
    \centering
   \includegraphics[width=1\linewidth]{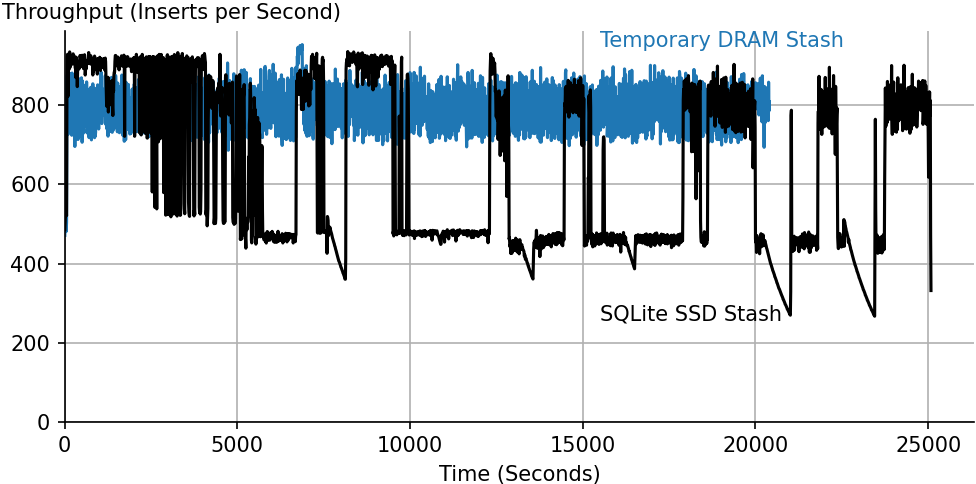}
    \caption{Throughput of an ephemeral DRAM stash when compared with a durable (SQLite) SSD stash ingesting 16 GB of data.}\label{fig:dramSSD}
\end{figure}
\begin{figure}[!ht]
    \centering
   \includegraphics[width=1\linewidth]{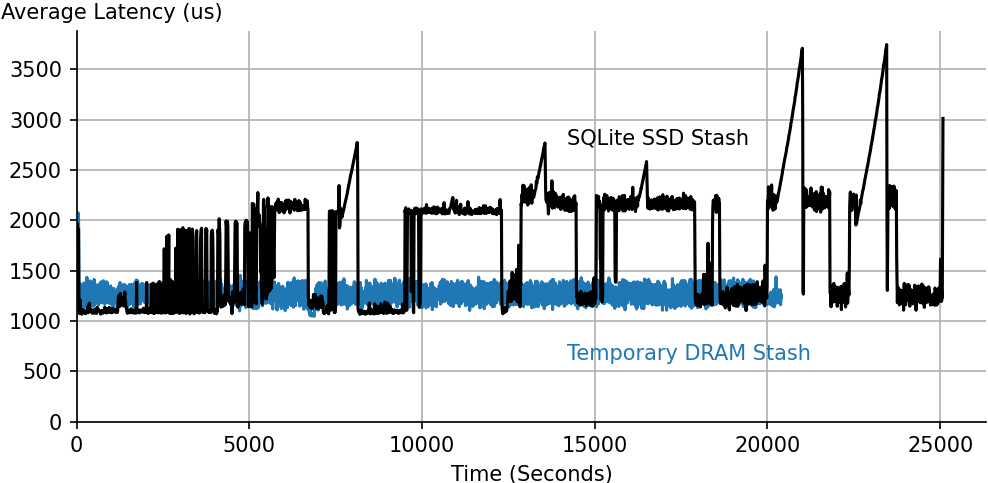}
    \caption{Latency of an ephemeral DRAM stash when compared with a durable (SQLite) SSD stash ingesting 16 GB of data.}\label{fig:dramSSDlatency}
\end{figure}

Figures~\ref{fig:dramSSD} and~\ref{fig:dramSSDlatency} shows the throughput and latency of an ephemeral DRAM stash and a durable SSD stash, respectively.
With each, 1 YCSB thread inserts 16 GB of data in the stash.
The DRAM stash inserts data at a rate of 850 transactions per second with CPU utilization of 15\%.
The SSD stash is configured with SQLite.
Its rate of data ingest varies from 480 to 900 transactions per second due to the DurabilityLag\footnote{These experiments were conducted using the eBay infrastructure.
With the HD stash experiments conducted using the CloudLab, the throughput varies from 0 to 120 transactions per second.}.
To explain further, FDB separates the client read path from its write path, processing reads using the stash (storage server) directly.
A stash uses MVCC to process read requests~\cite{zhou2021foundationdb}.
Its RateKeeper calculates the DurabilityLag, the difference between a stash's (i.e., a Storage Server's) durable version and the latest version known to the component that generates new transaction ids, i.e., the sequencer.
If the difference becomes too large then the stash (storage server) will not be able to process the read requests because their version is too old.
FDB's RateKeeper prevents this by slowing down the rate at which it commits transactions.
This results in a lower throughput and a higher latency with SSD.

With the SSD stash, the RateKeeper results in a higher latency once it limits the transaction rate, see Figure~\ref{fig:dramSSDlatency}.
The utilization of the SSD stash is almost 100\% in these experiments.
Hence, increasing the number of YCSB threads does not increase its throughput.
It is different with the DRAM stash as the resources are underutilized.
By increasing the number of YCSB threads to 10, 
the throughput increases to 10,000 inserts per second and the CPU becomes 100\% utilized\footnote{In these experiments, the utiliztaion of the log servers SSDs is 15\%.  The CPU of the DRAM stash becomes 100\% utilized because it must index the inserted key-value pairs and maintain the LRU linked list.}.

\subsubsection{eBay Production Graph}
\begin{figure}[ht!] 
    \centering
    \begin{minipage}[c]{0.48\textwidth} 
        \centering
        \begin{tabular}{||c || c c c||}
            \hline
            & 50$^{th}$ & 95$^{th}$ & 99$^{th}$ \\ [0.5ex]
            \hline\hline
            1Stash & 2.23 & 23.36 & 80.36 \\
            2Stash & 2.18 & 20.93 & 72.33 \\
            \hline
            \hline
            \% Improvement & 2\% & 10.4\% & 10\% \\
            \hline
            \hline
        \end{tabular}
        \captionof{table}{Latency with 1Stash and 2Stash, moderate workload.}\label{tbl:cmp}
    \end{minipage}%
    \hfill 
    \begin{minipage}[c]{0.48\textwidth} 
        \centering
        \includegraphics[width=1\linewidth]{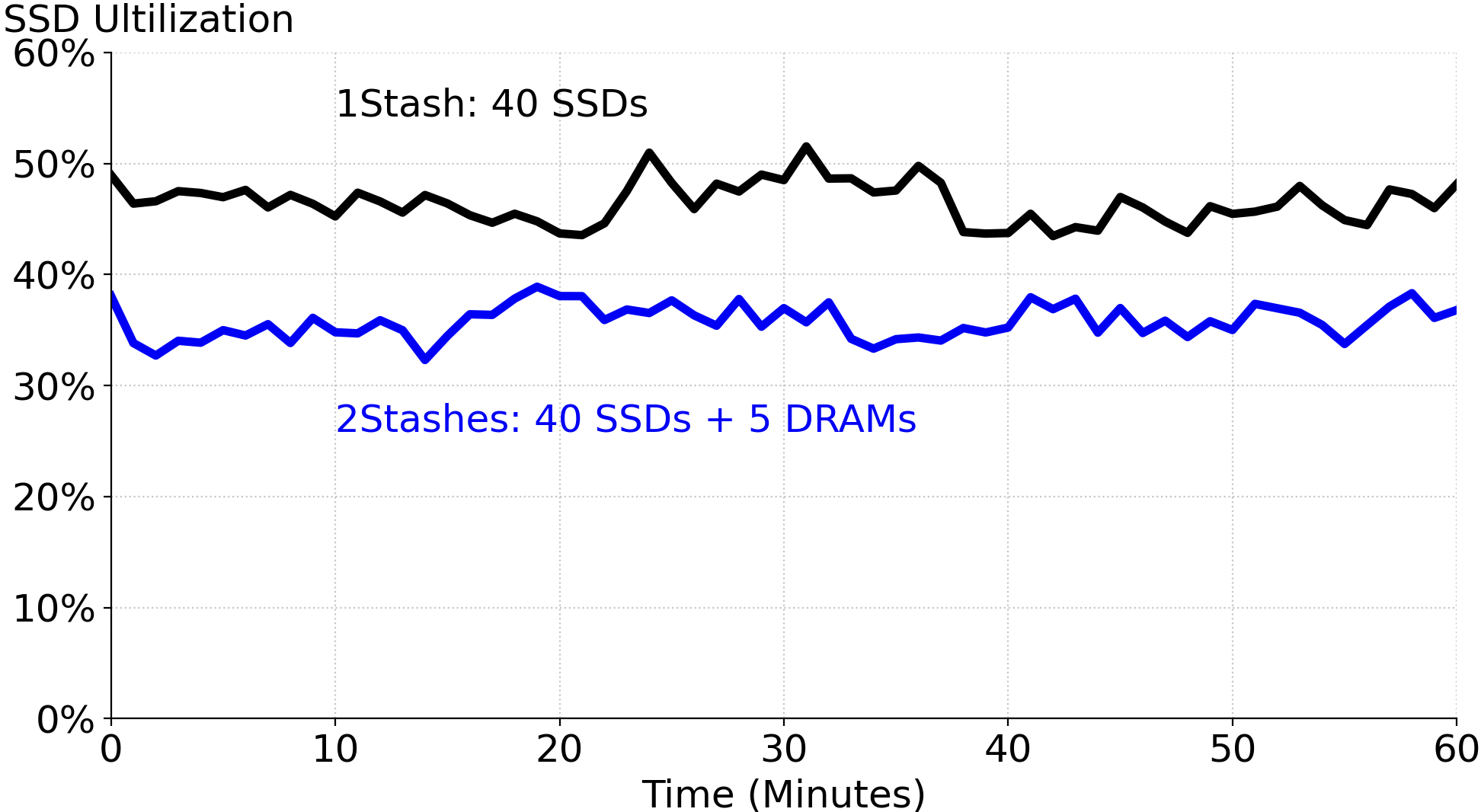}
        \captionof{figure}{SSD Utilization, moderate workload.}\label{fig:cmpSSDutil}
    \end{minipage}
\end{figure}



In the reported experiments, the cache hit rate is approximately 100\%.
Hence, approximately 45\% of the workload is processed using the query results.
This fraction uses 40 SSDs with 1Stash and 5 DRAMs with 2Stash.
With the light workload, 1Stash and 2Stash provide comparable performance.
With the moderate workload, Table~\ref{tbl:cmp} shows the average of 50$^{th}$, 95$^{th}$, and 99$^{th}$ latencies based on one hour of execution time.
While the 50$^{th}$ latency is comparable, 2Stash improves the 95$^{th}$ and 99$^{th}$ latencies by approximately 10\%. 
The DRAM stashes reduce the SSD utilization with 2Stash, see Figure~\ref{fig:cmpSSDutil}.
These freed SSD resources are used to expedite processing of the queries that do not observe a cache hit, i.e., 55\% of the workload.

With a moderate workload, the 50$^{th}$, 95$^{th}$ and 99$^{th}$ latencies of queries that observe a cache hit is comparable with both 1Stash (40 way parallelism using SSD) and 2Stash (5 way parallelism using DRAM).
The buffer pool manager of SQLite maintains the cache entries in its memory (DRAM), serving them as fast as the DRAM of the 2Stash.
This is at the expense of 55\% of the workload that reference disk pages for query processing.  
This expense is shown with a higher SSD utilization in Figure~\ref{fig:cmpSSDutil}.

\noindent{\em Cache Hit Latency.}\label{sec:cmp2configs}
During normal mode of operation and with a light system load, the 50$^{th}$ latency of 1Stash and 2Stash are comparable.
However, 1Stash reduces the 95$^{th}$ and 99$^{th}$ latencies, compare the start (x-axis=0) of Figures~\ref{fig:40ssdlat} and~\ref{fig:dramlat}.
This is because, with 2Stash, the load is not uniformly distributed across the 5 DRAMs.
Figure~\ref{fig:dramutil} (origin until 23:28) shows two of the stashes have a lower utilization than the other three.
Balancing the load across a few stashes with a light system load is non trivial.
The observed performance gap disappears with a moderate system load.
With this load, while the 50$^{th}$ latency remains unchanged, the 95$^{th}$ latency of both configurations increases to 5.3 ms, and their 99$^{th}$ latency increases to 15 ms.

\subsection{Failure Mode of Operation}
With failures, the number of impacted stashes in an epoch is $x$=1 with the microbenchmark and $x$=2 with the eBay production workload. 

\subsubsection{Summary of Results:}
DiStash is able to tolerate failure of $R-1$ stash failures where $R$ is the degree of data replication with the stash pool.  Its DD re-constructs replicas of those shards with fewer than $R$ replicas aggressively.
DiStash uses these same replicas for query processing.
This causes the load on the servers hosting these replicas to increase.
Depending on the system load, there may be a noticeable increase in the 50$^{th}$, 95$^{th}$, 99$^{th}$ latencies.

Insertion of one or two stashes (at the end of an epoch) causes DiStash, i.e., its DD, to reorganize data across stashes.  With a few stashes (5 DRAMs of 2Stash), this increases the 95$^{th}$ and 99$^{th}$ latencies temporarily with a light system load.  With a large number of stashes (40 SSDs of either 1Stash or 2Stash), there is no noticeable change in latency with a light system load.

\subsubsection{Microbenchmark: 
 Ephemeral Stash Failure}
\begin{figure}[ht]
    \centering
    \begin{subfigure}[b]{0.5\textwidth}
        \centering
        \includegraphics[width=\textwidth]{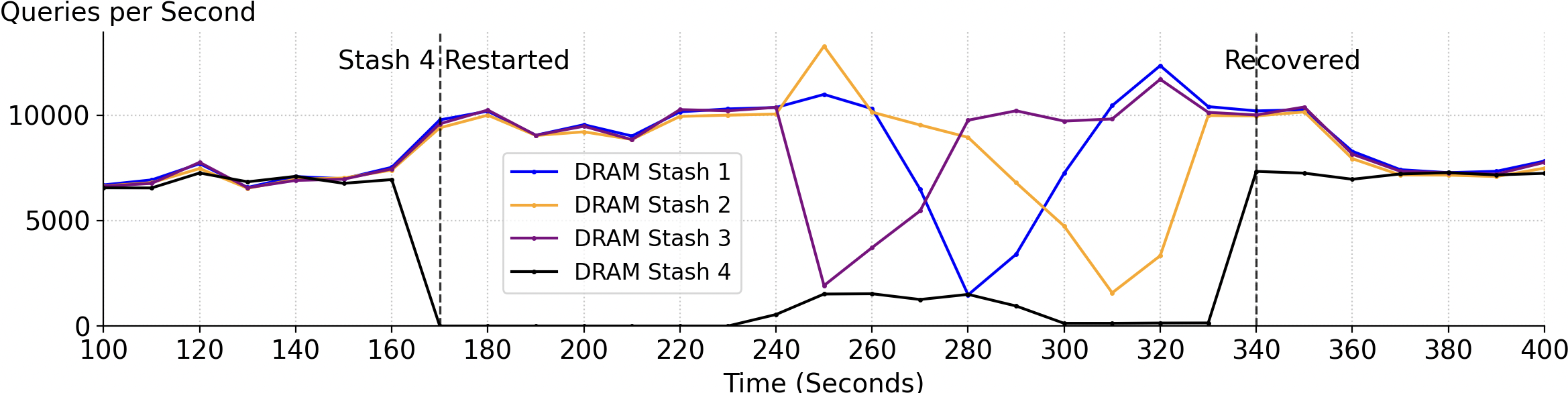}
        \caption{Rate of query processing per DRAM stash.}
        \label{fig:qrate}
    \end{subfigure}
    \hfill
    \begin{subfigure}[b]{0.5\textwidth}
        \centering
        \includegraphics[width=\textwidth]{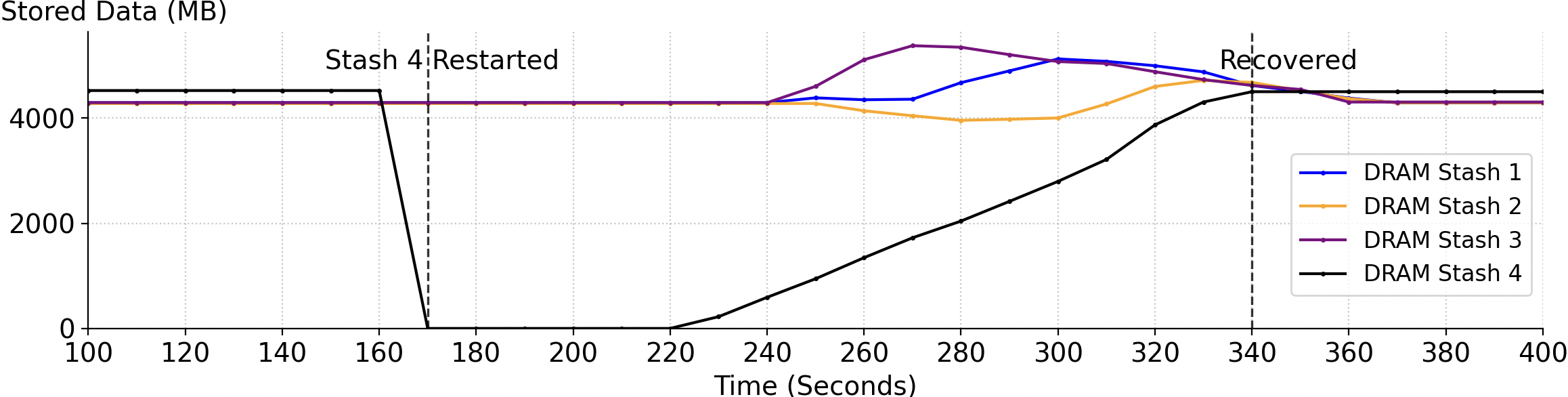}
        \caption{Data Size per DRAM stash.}
        \label{fig:datasize}
    \end{subfigure}
    \hfill
    \begin{subfigure}[b]{0.5\textwidth}
        \centering
        \includegraphics[width=\textwidth]{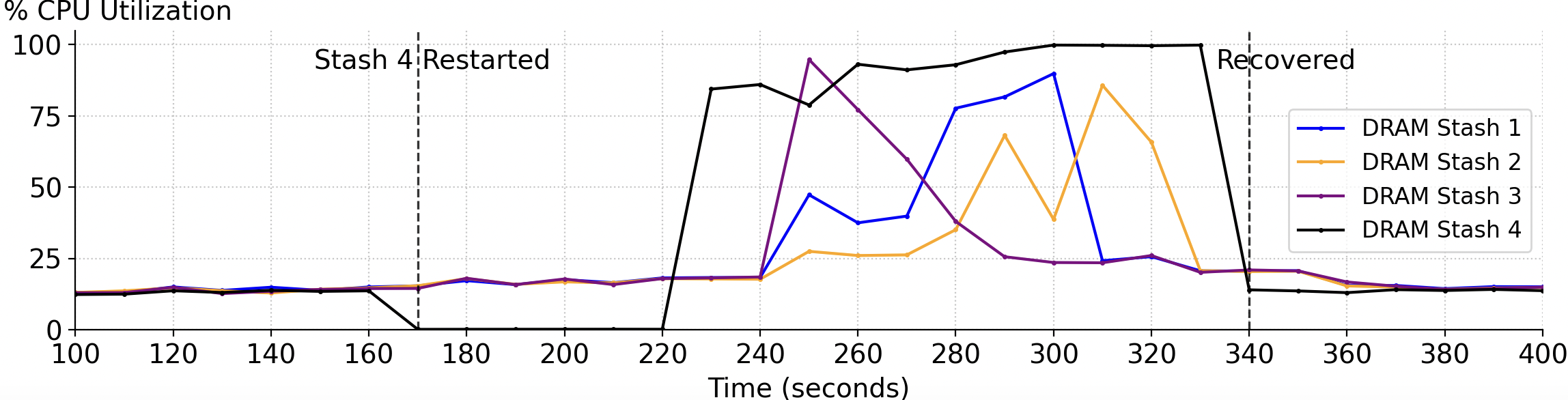}
        \caption{CPU Utilization of each DRAM stash.}
        \label{fig:failutil}
    \end{subfigure}
    \caption{Failure and recovery of an ephemeral volatile stash.}
    \label{fig:failuremicro}
\end{figure}

After a failure, DiStash recovers an ephemeral DRAM stash by incorporating it as a new instance of the stash.
This is because its content has been lost.
With a degree of replication of two or higher, a replica of this content is available and DiStash uses a replica to recover the content of the failed stash.

Figure~\ref{fig:failuremicro} shows the failure and recovery of an ephemeral DRAM stash.  
This experiment is across 3 data centers.
A DRAM stash pool consisting of 8 stashes is spread across two data centers, 4 ephemeral DRAM stashes per data center.
A third data center contains a log server.
Each stash pool has a degree of replication of 3. 
Total database size is 5 GB and generated using YCSB.
The foreground system load is generated by a YCSB client with 10 concurrent threads.
Each thread reads one key-value pair selected randomly using a uniform distribution.

A time 170, we restart Stash 4 in Data Center 1 (DC1).
Team Collection of the Data Distributor, DDTC, is not aware whether Stash 4 has failed or is partitioned from the network.
Hence, it waits for 60 seconds for this stash to rejoin the network.
During this time, i.e., from time 170 to 230, the replicas of Stash 4's shards process its load.
This is shown with an increase in the rate of query processing for Stashes 1-3 at time 170 in Figure~\ref{fig:qrate}.

DiStash recovers the failed non-volatile stash by introducing it as a new one.  
At time 230, DDTC incorporates this new instance and starts to assign shards to it, see Figure~\ref{fig:datasize}.
It is unable to use this instance sooner because the system configuration required it to wait for 60 seconds for the old instance with its data to return.
At the same time, both the DDTC and the DDQ (DD components) start to aggressively re-construct the missing replica of the lost shards due to the failure of Stash 4.
This is shown by an increase in the amount of data stored on Stashes 1 to 3, see Figure~\ref{fig:datasize}.
The missing replica is constructed on the remaining stashes one at a time, starting with Stash 3, followed by 1, and 2.
This causes the CPU of each stash to become fully utilized one after another, see Figure~\ref{fig:failutil}.
It is aggressive by giving a higher priority to constructing the missing replica of Stash 4 instead of processing requests.
This causes the stash's rate of query processing to decrease significantly, see Figure~\ref{fig:qrate}.

Note that the CPU of the new stash (failed Stash 4 that was restarted) is utilized fully from time 230 to time 330.
During this time, DD\footnote{The shard tracker is responsible for splitting and merging shards (ranges).  Migrations of Figure~\ref{fig:failuremicro} are at the granularity of ranges.} (DDTC and DDQ) is balancing the load across the stashes by migrating shards to the new stash.  
After time 330, the recovery of the restarted DRAM stash is complete.
Note that the other stashes have an elevated level of query processing rate from Time 340 to 360.
DD is balancing the load across them and is deleting their redundant copies of shards during this period.

\subsubsection{eBay Production Graph}
We evaluated the impact of stash failures on 1Stash and 2Stash cluster configurations. This analysis specifically reports on queries that observe a cache hit.
We repeat an epoch to verify the observed trends are reproducible.

\begin{figure}[!ht]
    \centering
   \includegraphics[width=1\linewidth]{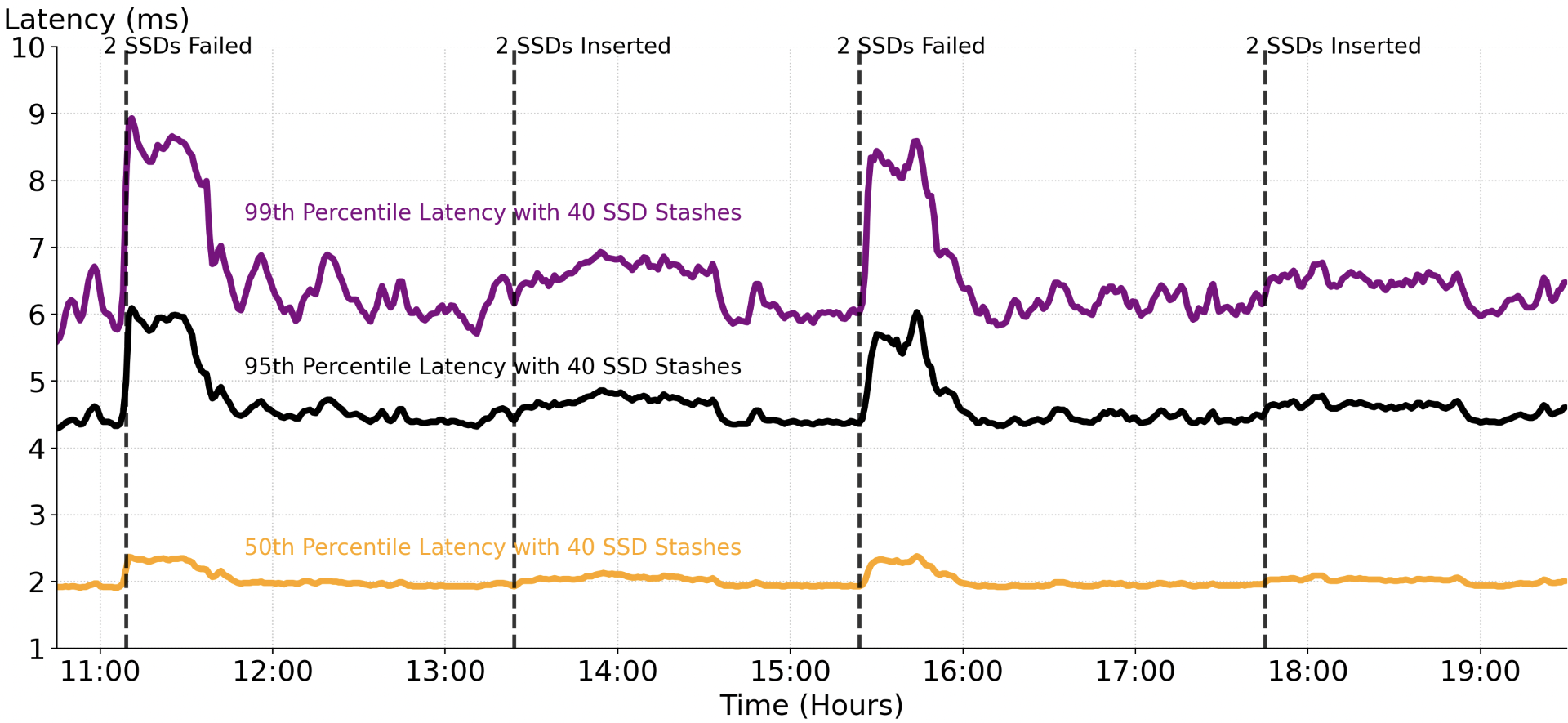}    \caption{Query result lookup (cache hit) latency of 1Stash with 40 SSDs with light workload.  There are two epochs.  Each starts with the failure of 2 stashes followed by their re-introduction in the cluster.  The first epoch starts at 11:09 and ends at 13:24.
    The second starts at 15:24 and ends at 17:45.}\label{fig:40ssdlat}
\end{figure}

\begin{figure}[!ht]
    \centering
   \includegraphics[width=1\linewidth]{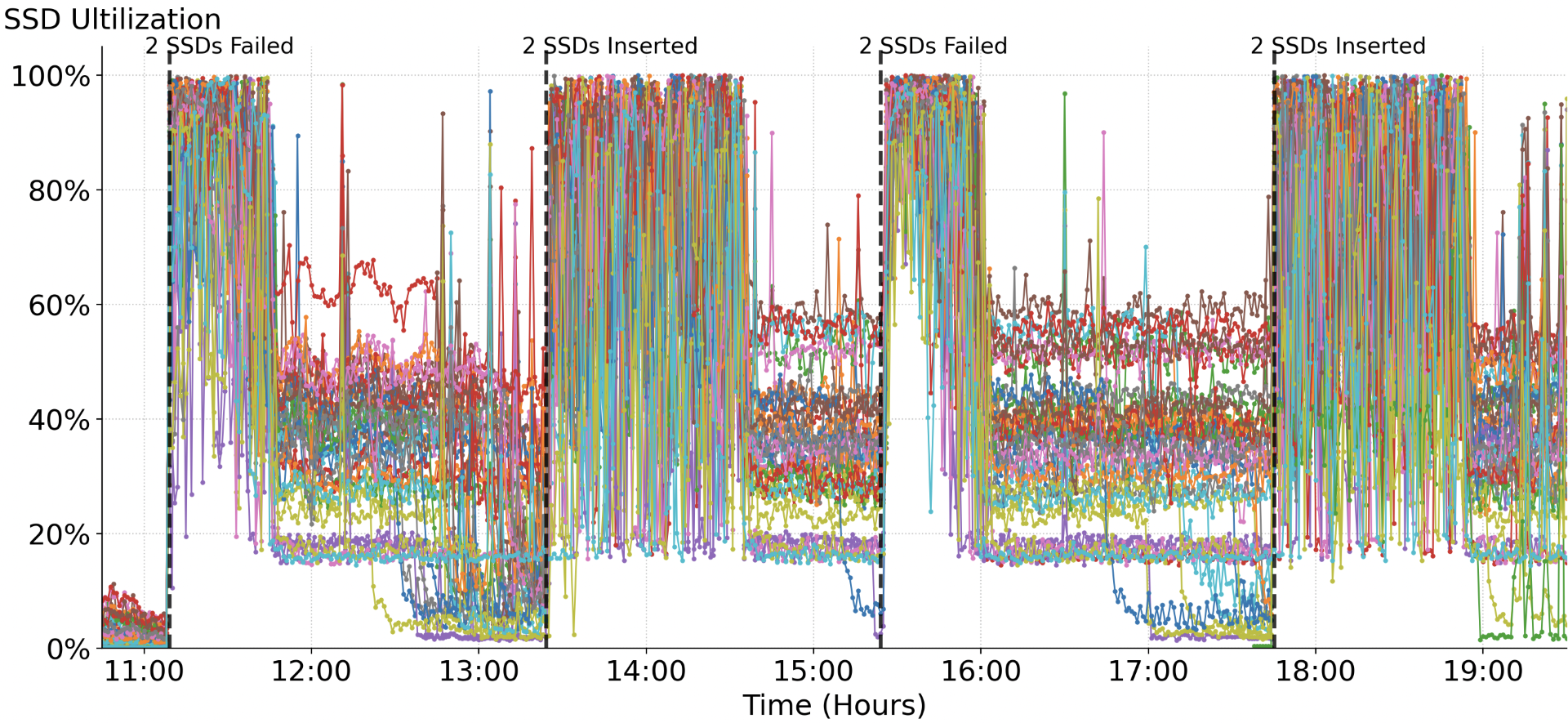}
    \caption{Utilization of 40 SSDs with epochs of Figure~\ref{fig:40ssdlat}.}\label{fig:ssdutil}
\end{figure}

Figure~\ref{fig:40ssdlat} shows the 50$^{th}$, 95$^{th}$ and 99$^{th}$ latencies with 1Stash.
The failure of 2 SSDs causes the DD to replicate the shards with missing replicas, i.e., those with either 1 or 2 surviving replicas.
These shards correspond to both the graph data and the query result data.
Their duplication across the remaining 38 SSDs causes the disk to become 100\% utilized\footnote{DD strives to ensure the same each stash performs its fair share of work.}, see Figure~\ref{fig:ssdutil}.

Once we insert 2 SSDs back into the 1Stash configuration, the DD starts to migrate shards to them in order to utilize their available bandwidth.
This causes the utilization of some SSDs to spike to 100\%.
The migration is not aggressive and has minimal impact on the 95$^{th}$ and 99$^{th}$ latencies.
This re-balancing requires a long time because it involves shards of both the graph data and the query result data.

\begin{figure}[!ht]
   \includegraphics[width=1\linewidth]{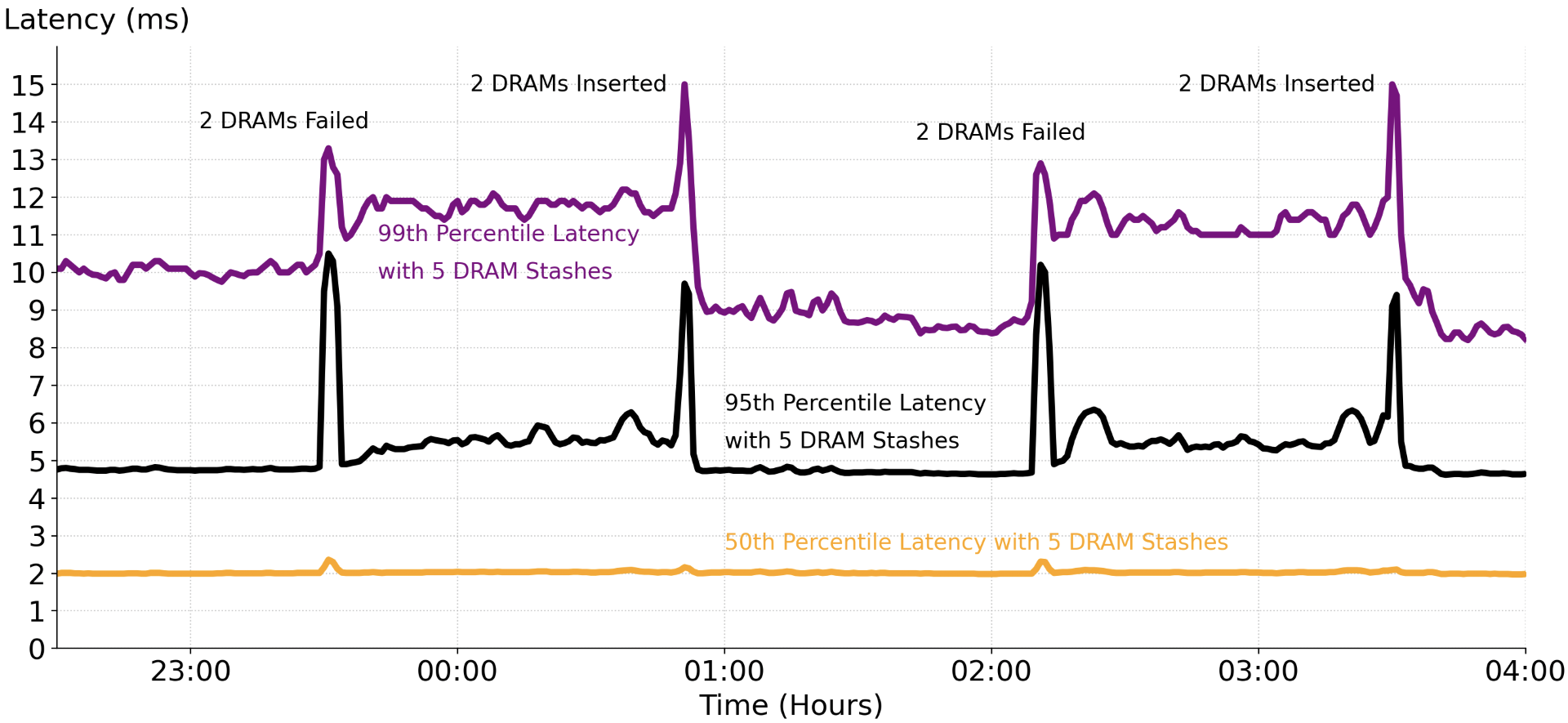}
    \caption{Query result lookup (cache hit) latency of 2Stash, 40 SSD + 5 DRAM with light workload.  Similar to Figure~\ref{fig:40ssdlat}, there are two epochs.  The first starts at 23:28 and ends at 00:40.  The second starts at 2:09 and ends at 3:29.}\label{fig:dramlat}
\end{figure}

\begin{figure}[!ht]
    \centering
   \includegraphics[width=1\linewidth]{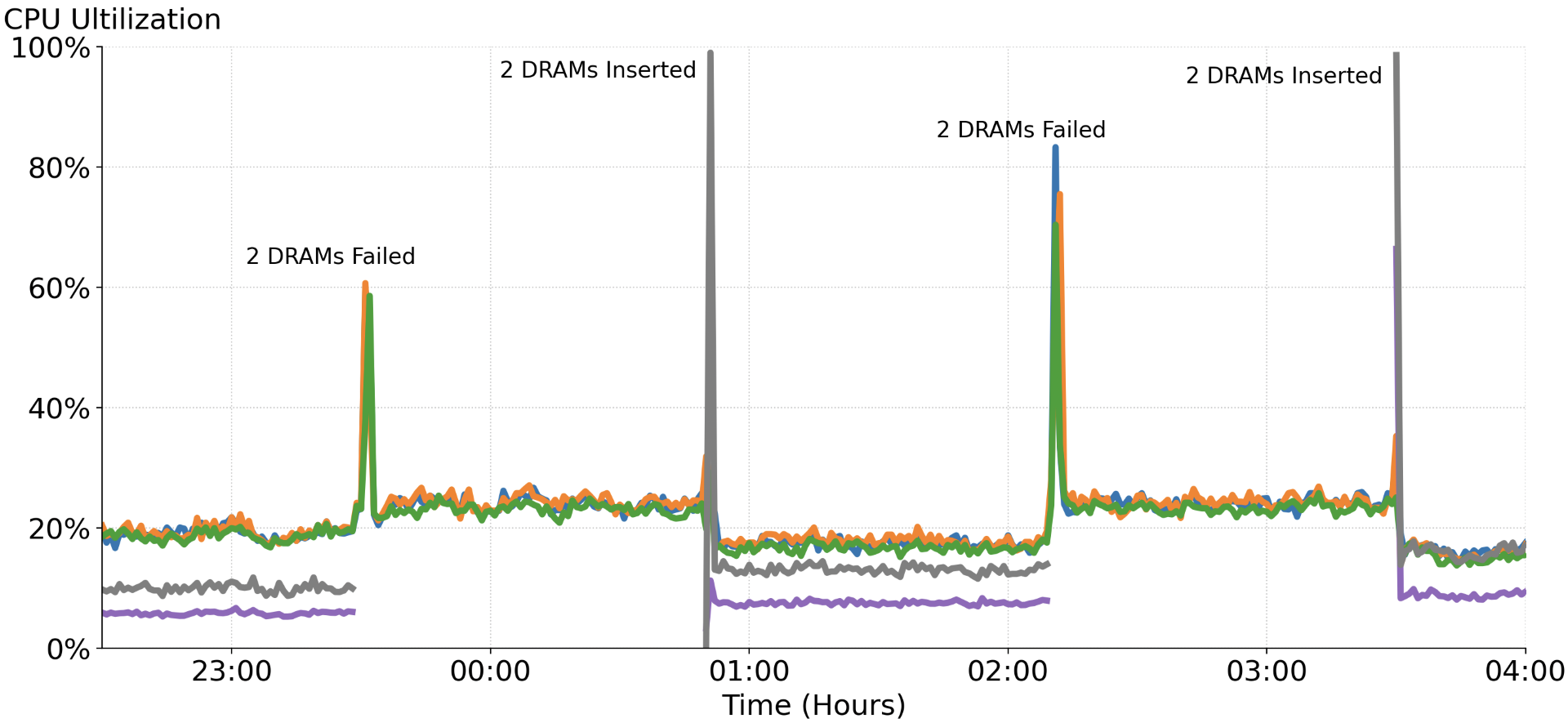}
    \caption{CPU utilization of 5 DRAMs with epochs of Figure~\ref{fig:dramlat}.}\label{fig:dramutil}
\end{figure}

Figure~\ref{fig:dramlat} shows the 50$^{th}$, 95$^{th}$ and 99$^{th}$ latencies with 2Stash.
Failure of 2 DRAM stashes results in aggressive replication of those shards with 1 or 2 missing replicas.
These are depicted as spikes at time 23:28 and 2:09.
Replication increases the CPU utilization of the remaining 3 DRAMs, see Figure~\ref{fig:dramutil}.
The impacted shards correspond only to the query results\footnote{The graph data is separate and stored across 40 SSDs.}.
They are compact in size.
Hence, the duration of aggressive replication is shorter when compared with 1Stash.
This is highlighted by the shorter duration of the spikes in Figure~\ref{fig:dramlat} compared to those in Figure~\ref{fig:40ssdlat}.

With 2Stash, after aggressive replication, the 95$^{th}$ and 99$^{th}$ latencies remain elevated because only 3 stashes are processing requests.
This results in a higher CPU utilization during each epoch, see Figure~\ref{fig:dramutil}.

Once we insert the 2 DRAM stashes, the latency and CPU utilization spike as DiStash starts to migrate data to them.
This re-balancing decreases the 95$^{th}$ and 99$^{th}$ latencies quickly as 5 DRAM stashes start to process requests (instead of 3). 

\noindent{\em Cache Hit Latency.} With 2 stash failures in the caching layer, the duration of time to replicate shards with 1 or 2 replicas is longer with 1Stash.
By co-locating graph data with query results on the SSDs, 1Stash must reconstruct a larger amount of data when compared with 2Stash.
1Stash continues to provide a lower 95$^{th}$ and 99$^{th}$ latency because, after the two failures, it uses 38 SSDs in parallel instead of 3 DRAMs with 2Stash.

\section{Related Work}\label{sec:related}
To the best of our knowledge, DiStash is the first transactional storage manager to support $h$ pools of stashes across multiple data centers with a different degree of data replication for each pool.  It enables its application to use a transaction to migrate or copy one or more key-value pairs across one or more stashes in a pool or different stash pools.  

As detailed in Section~\ref{sec:motivate}, DiStash may be used to implement cache-augmented database management systems~\cite{writeback2019,mercury2012,ebay2023,scalingmemcache2013,onehoparXiv}, load balancing techniques using a front-end cache~\cite{smallcache,netcache2017}, and hierarchical storage structures~\cite{flashstore2010,hierarchical1995,nhc2021}.
However, these systems and techniques cannot be used to implement DiStash.
For example, with a cache augmented solution, one may use the IQ framework~\cite{iq2014,trig2017} to maintain consistency of data with a query result cache.
However, this framework is specific to the write-around policy.
It does not support multiple stash pools or the write-through policy. 
Fundamentally, DiStash provides consistency using one transaction to manage data across multiple stash pools.
This concept is absent from other systems.

In~\cite{config2018,config2018arXiv}, we present a simple algorithm to evaluate
design tradeoffs in the context of a memory hierarchy for a
key-value store.  This algorithm answers questions such as, given a fixed budget and a workload, which storage media should constitute a memory hierarchy?  And, whether the data should be replicated or tiered.
This offline algorithm may be used to size the different stash pools of DiStash for different workloads and applications.

\section{Conclusions and Future Work}\label{sec:conc}
DiStash is a transactional storage manager for $h$ pools of stashes.
It enables an application to use a single transaction to read and write copies of key-value pairs in a stash pool and across different stash pools.
A stash pool might be either ephemeral or durable.
DiStash may be used to implement cache-augmented database management systems, novel load-balancing techniques, hierarchical storage structures, and a hybrid of these.  We presented performance numbers from DiStash 
using eBay's production workload.
This evaluation highlighted the flexibility of DiStash to use durable store for both graph data and query results (cache entries), and use of durable store for graph data and ephemeral store for cache entries.
We showed DiStash is able to tolerate failure of stashes in either durable or ephemeral storage.
Many configurations of DiStash are possible, including those based on DRAM alone, e.g., RAMCloud~\cite{ramcloud2015}.

This workshop paper identifies many interesting future research directions.
One is how to balance load with a few stashes and a light system load, see discussions of Figure~\ref{fig:dramutil}.
This requires a global analysis of how FoundationDB (FDB) balances load with a light, moderate, and heavy system loads.  At the time of this writing, FDB balances load effectively with a moderate and a heavy system load.  In addition, we plan to integrate the configuration planner of~\cite{config2018,config2018arXiv} with the alternative workloads to size DiStash.
This will include a comprehensive analysis of workloads with write transactions and the hybrid architecture of Figure~\ref{fig:stashpools}. 
Moreover, we plan to extend DiStash with an ephemeral storage finetuned for NVM~\cite{patil2025nvm,wang2025boosting} including NVDIMM~\cite{lee2025boosting,clouzet2024h2m,lee2020nvdimm}.

\balance

\begin{acks}
This research was a collaborative effort between eBay and the
University of Southern California, made possible through support
from eBay’s eRUPT program.
We gratefully acknowledge CloudBank~\cite{cloudbank2021} and CloudLab~\cite{emulab} for the use of their resources to enable some of the experimental results presented in this article.
In addition to eRUPT,
Shahram Ghandeharizadeh was supported in part by the NSF grant CMMI-2425754.
\end{acks}

\bibliographystyle{ACM-Reference-Format}
\bibliography{refs}

\end{document}